\documentclass[aps,prl,twocolumn,superscriptaddress,amsfonts,amsmath,amssymb]{revtex4-1}

\usepackage{graphicx}
\usepackage{hyperref}
\begin{document}

%\preprint{}

\title{The footprint of atmospheric turbulence in power grid frequency measurements}


\author{H. Haehne}
\author{J. Schottler}
\author{M. Waechter}
\author{J. Peinke}

\affiliation{Institute of Physics and ForWind, Carl von Ossietzky University Oldenburg, 26111 Oldenburg, Germany}

\author{O. Kamps}

\affiliation{Center for Nonlinear Science, University of Muenster, 48149 Muenster, Germany}

%\thanks{}


%\date{\today}

\begin{abstract}
Fluctuating wind energy makes a stable grid operation challenging. Due to the direct contact with atmospheric turbulence, intermittent short-term {variations} in the wind speed are converted to power fluctuations that cause transient imbalances in the grid. {We investigate the impact of wind energy feed-in on short-term fluctuations in the frequency of the public power grid, which we have measured in our local distribution grid.} By conditioning on wind power production data, provided by the ENTSO-E transparency platform, we demonstrate that wind energy feed-in has a measurable effect on frequency {increment} statistics for short time scales \mbox{($<$ 1 sec)} that are below the activation time of frequency control. Our results {are in accordance} with previous numerical studies of self-organized synchronization in power grids under intermittent perturbation and rise new challenges for a stable operation of future power grids fed by a high share of renewable generation.
\end{abstract}

%\pacs{}
%\keywords{}

\maketitle

\section{Introduction}

Wind energy is one of the core elements of renewable power production with increasing feed-in to the central European power grid: In 2016, an installed capacity of 153.7 GW in the EU generated almost 300 TWh and covered 10.4 \% of the EU electricity demand \cite{windeurope2017}. {In the first days of 2018, even more than 20\% (60\%) of the EU (German) daily electricity demand was covered \cite{dailywind}.}

A stable and reliable supply with electrical power is essential for both, society and economy. The power grid frequency reflects the transient ratio of production to demand in the grid and thus serves as an instantaneous and locally inferable stability parameter. {Mismatch of production and consumption causes frequency deviations from the nominal frequency} \cite{kundur1994power}. {Load frequency} control of the grid operator restores the frequency after perturbations: The fastest control (``primary control'') sets in seconds after a deviation from the nominal frequency to stabilize, but not yet restore, the frequency. Restoration is achieved by secondary control which operates on time scales of 30 seconds and {beyond} \cite{operation}. 

Wind energy feed-in is known to be highly volatile. Fluctuations of a process $x(t)$ on a time scale $\tau$ are often characterized by means of increments {\mbox{$\Delta_\tau x := x(t) - x(t+ \tau)$}}. Traditional analysis and prediction of wind speed considers variations in 15 minutes and longer \cite{albadi2010overview, anvari2016short}. However, recent findings in the analysis of short-term increments of renewable power generation reveal strongly non-Gaussian fluctuations even on scales of one second \cite{milan2013turbulent, anvari2016short}. But, where does this short-term behavior result from?

The atmospheric boundary layer is known to be non-stationary and turbulent \cite{boettcher2003statistics, baile2010spatial, calif2013multifractal}. {Turbulent flows show scale-dependent increment statistics: In an hierarchical cascade process, kinetic energy is transferred from large- to small-scale structures \cite{renner2001experimental}.} Specifically, this implies pronounced tails in short-term increment statistics; an effect termed {\it intermittency} in turbulence research \cite{frisch1995turbulence}. Due to the intermittent increment statistics, severe wind-speed fluctuations are much more likely than expected from a normal distribution. 

A wind turbine transforms the kinetic energy of the wind to electric power. Even though {\it ac-dc-ac} converters decouple wind speed from power output dynamics, turbine controllers maximize the power output and follow wind speed variations \cite{milan2013turbulent}. Hence, atmospheric fluctuations even within a second propagate into the power output of wind farms and are fed to the grid. In fact, intermittency is found in production time series of wind (and also photovoltaic) power plants \cite{anvari2016short, milan2013turbulent}. Consequently, power quality is a major challenge for the grid integration of renewable generators \cite{liang2017emerging}. 

The transient short-term reaction of power grids to perturbations has attracted great attention in theoretical physics: Simple models of high-voltage ac-grids correspond to complex networks of Kuramoto-like, phase-coupled oscillators \cite{filatrella2008analysis, schmietendorf2014self}. Such models have been used to analyze aspects of synchronization \cite{rohden2012self} and the interplay of stability and topology \cite{menck2014dead}, as well as relaxation after singular \cite{kettemann2016delocalization} and stochastic \cite{schafer2017escape, schmietendorf2017impact} perturbations. The impact of intermittent feed-in on power grids has been addressed in \cite{schmietendorf2017impact, auer2017stability}: Numerical results indicate that intermittency propagates in a power grid and affects the frequency increment distributions of nodes distant to the feed-in. Stochastic models for non-Gaussian frequency fluctuations are presented in \cite{schaefer2017nongaussian}. However, none of the prior results relates the intermittent feed-in to transient stochastic properties of the grid frequency.

\begin{figure}
\centering
\includegraphics[scale=0.6]{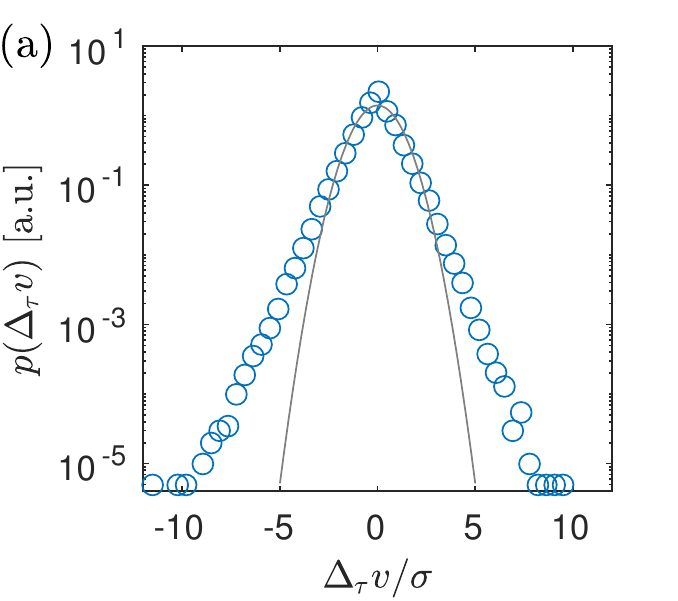}\includegraphics[scale=0.6]{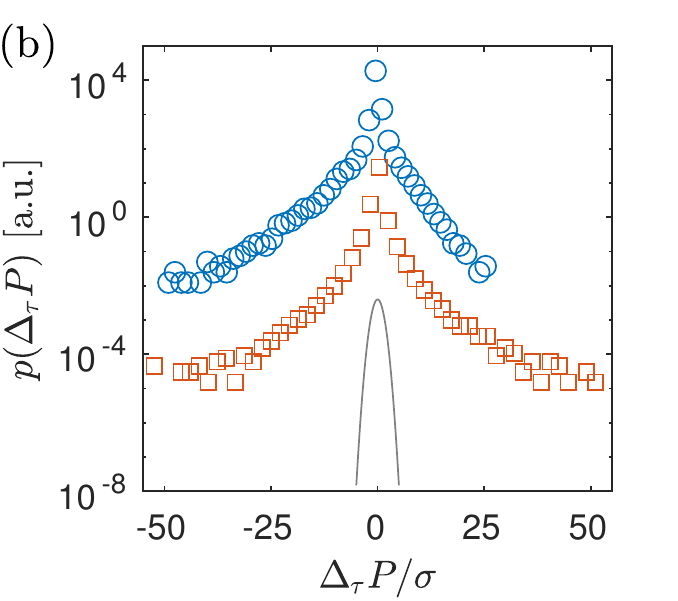}
\caption{{\bf Atmospheric intermittency is preserved in power time series of a single wind turbine and also in the average output of a farm of twelve turbines.} {\it (a)} Distribution of wind speed increments {$\Delta_\tau v = v(t) - v(t+\tau)$} for $\tau = 1$ sec. Due to the turbulent conditions in the atmosphere, the increment distribution shows large deviations from the normal distribution (gray). {\it (b)} Distribution of power increments $\Delta_\tau P$ for $\tau = 1$ sec of a single turbine (blue) and of the average power of twelve turbines (orange). The deviations from the normal distribution are even more pronounced than for the wind speed increments and do not average out. {The increment PDF for the single turbine is not symmetric which, to our interpretation, results from operation close to the rated power $P_r \approx 2$ MW.} All increments are given in units of their respective standard deviations; these are $\sigma(\Delta_\tau v) = 0.29$ m/s, {$\sigma(\Delta_\tau P) = 0.0067$ MW} for the single turbine, and {$\sigma(\Delta_\tau P) = 0.0292$ MW} for the average of twelve turbines. {PDFs are vertically shifted for convenience of presentation.} {Figures are similar to those in \cite{anvari2016short} and were produced from a freely available \cite{downloadData} data set of 1 Hz-recordings of twelve onshore turbines during one month; kindly provided by WPD Windmanager GmbH, Bremen, Germany. A detailed stochastic analysis can be found in \cite{milan2013turbulent} and \cite{anvari2016short}.} }
\label{fig:intermittency}
\end{figure}

It is often believed that intermittency vanishes when power time series of many turbines are averaged. To support this hypothesis, usually the central limit theorem is referred to. Velocity time series $v(t)$ are, however, highly correlated \cite{muzy2010intermittency} and so are the resulting power time series \cite{anvari2016short}. {The lacking statistical independence} makes this theorem inapplicable. Intermittency is, in fact, observed in power outputs of entire wind farms \cite{anvari2016short} and withstands even country-wide averaging \cite{kamps2014characterizing}. We show, as an example, the increment distribution for the mean power output of twelve turbines in comparison to the one of a single turbine in Fig.\ \ref{fig:intermittency}. But, what exactly is the impact of wind power feed-in on the grid frequency?

In this Letter, we complement the previous numeric and analytic work and show that the feed-in of intermittent wind power has a measurable effect on the increment statistics of the frequency measured in the distribution grid. Instead of focusing on the frequency response to singular, large-deviation events, as for example in \cite{lauby2014frequency}, we use the full statistical information encoded in the increment statistics of the grid frequency and focus on time scales that lie below the activation of primary control.  \\

This paper continues as follows: We first introduce our measurement and data processing techniques. Subsequently, we show our stochastic analysis and its interpretation. Finally, we conclude and give an outlook on further research.

\section{Methods}
\label{sec:methods}

Publicly available measurements of the frequency of the public power grid have, to our knowledge, only a time resolution of {1 second or above}. We, however, want to observe the self-organized, transient behavior of the grid and thus need a higher time resolution.

We took 10 kHz voltage samplings $u(t)$ of a single phase of the distribution grid in our lab in Oldenburg, {northern} Germany, from November 8, 2016, till March 23, 2017. Subsequently, we applied the method of {\it Instantaneous Frequency} (IF) \cite{boashash1992estimating} to estimate the frequency time series $f(t)$ from the sinusoidal voltage signal $u(t)$.

{The IF reveals the dominant frequency component at each time instant $t$ and is thus suited for signals composed of one major frequency component. The method makes use of the fact that real-valued signals, such as the voltage signal $u(t)$, have conjugate symmetric Fourier representations, $\mathcal{F}[u](-\omega) = \mathcal{F}[u](\omega)^*$. Here, $\mathcal{F}$ denotes Fourier transform. The complex-valued analytic signal $z(t)$ is the inverse Fourier transform of the positive frequencies $\omega > 0$. Discarding the redundant negative frequency components makes the IF accessible. It is defined as the time derivative of the phase $\Phi(t)$ of the analytic signal $z(t)$:

\begin{equation}
f(t) = \frac{1}{2\pi}\frac{\mathrm{d}}{\mathrm{d}t}\Phi(t) = \frac{1}{2\pi}\frac{\mathrm{d}}{\mathrm{d}t}\mathrm{arg}(z(t)). \label{eq:if-def}
\end{equation}
In practice, $z(t)$ is obtained from the Hilbert transform $\mathcal{H}[u](t)$ of the original signal: \mbox{$z(t) := u(t) + i\mathcal{H}[u](t)$}. The Hilbert transform can be obtained from \mbox{$\mathcal{H}[u](t) = (u * 1/\pi t' ) (t),$} where ``$*$'' denotes convolution.}

\begin{figure}
\centering
\includegraphics[scale=0.6]{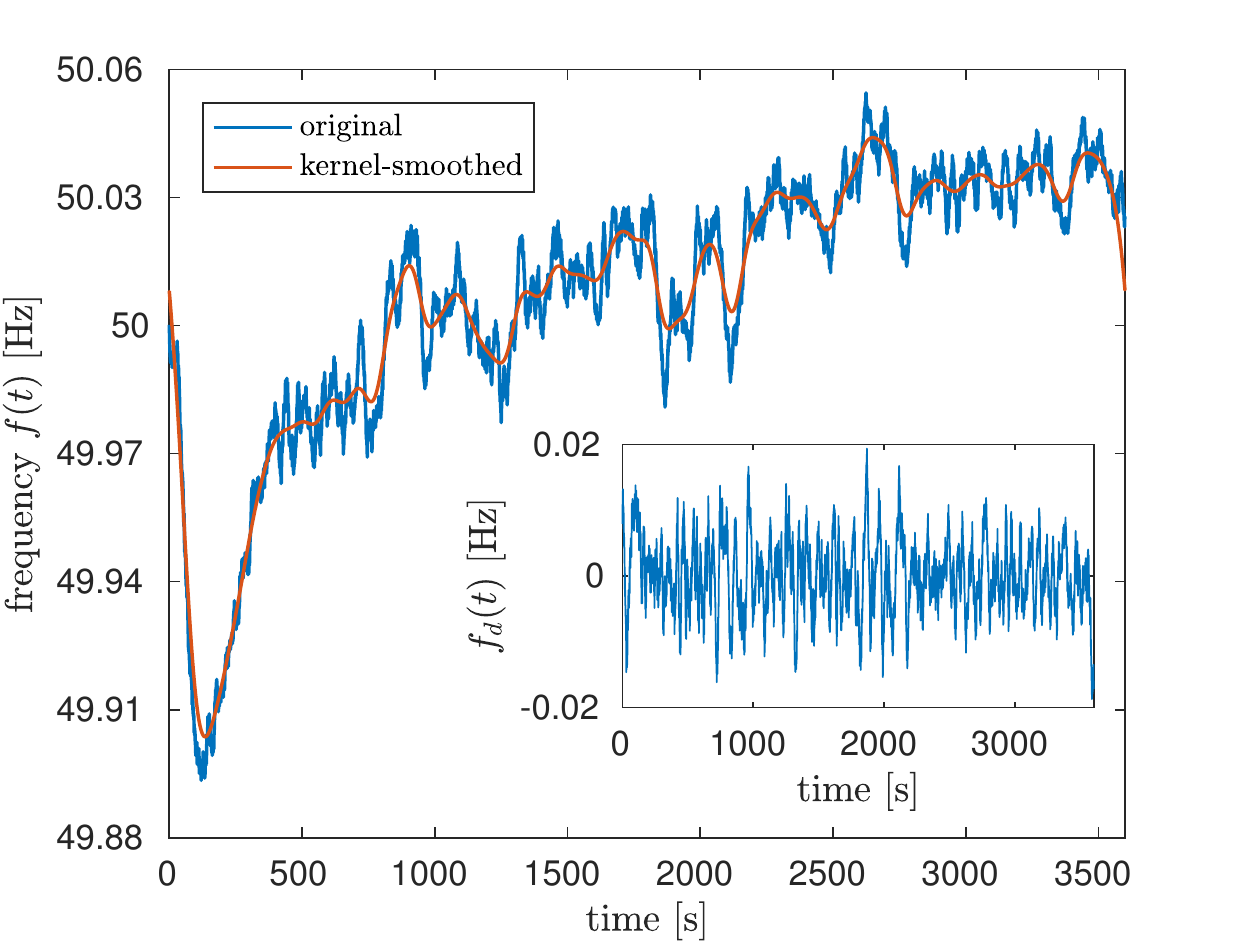}
\caption{{\bf Detrending isolates the short-term behavior of the signal.} Original frequency signal $f(t)$ (blue) and kernel-smoothed signal $f_{ks}(t)$ (orange) in one example hour. The kernel standard deviation is $\sigma =$ 30 s. {\it Inset}: Detrended signal $f_d(t)$.}
\label{fig:detr}
\end{figure}

To estimate the derivative in Eq.\ \eqref{eq:if-def} numerically, the phase $\Phi(t)$ was calculated for every time step in the voltage signal. Subsequently, the time derivative was estimated by linear fits of $\Phi(t)$ in disjoint blocks of 2000 samples. This procedure gives a frequency time series $f(t)$ with a time resolution of 200 ms. The $2\sigma$-confidence bounds of the linear fits are, in average, of size $\pm$ 1 mHz. We show one example hour from our measurements in Fig.\ \ref{fig:detr}.\\

Frequency measurements of the public power grid are influenced by many factors that overlay the influence of renewable generation. The signal shows severe deviations from the nominal frequency each full and half hour caused by power trading. Further, it is influenced by long-term correlations in demand and production. Thus, we applied {\it Kernel Detrending} to isolate the short-term behavior of the frequency signal $f(t)$: A kernel-smoothed signal $f_{ks}(t)$ was subtracted from the original to obtain the detrended signal $f_d(t) = f(t) - f_{ks}(t)$. We used a 30-seconds Gaussian kernel. The kernel-smoothed signal is obtained from convolving the original signal with a Gaussian curve $g_\sigma(t)$ with standard deviation $\sigma =$ 30 s and zero mean:

\begin{align}
f_d(t) = f(t) - f_{ks}(t) = f(t) - (f * g_\sigma)(t).
\end{align}
{Again, ``$*$'' denotes convolution.} We illustrate the detrending in Fig. \ref{fig:detr}. In the following, we use the detrended data and drop the index $d$.

\section{Results}
\label{sec:results}

\begin{figure}
\centering
\includegraphics[scale=0.6]{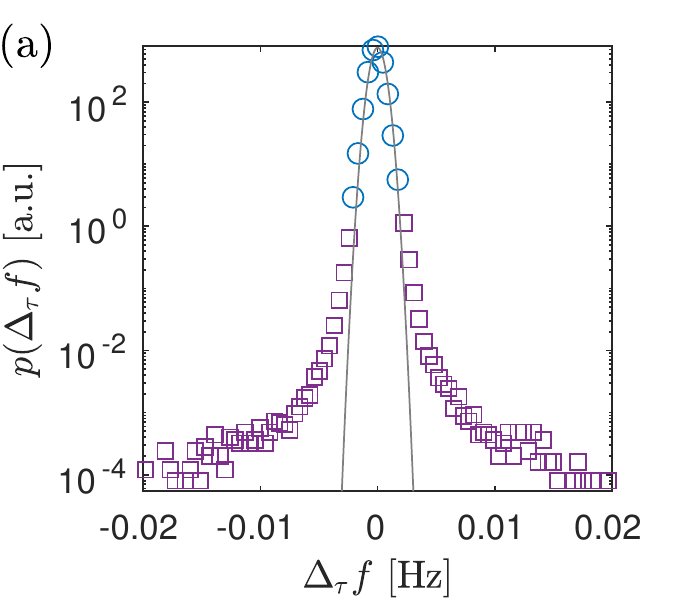}\includegraphics[scale=0.6]{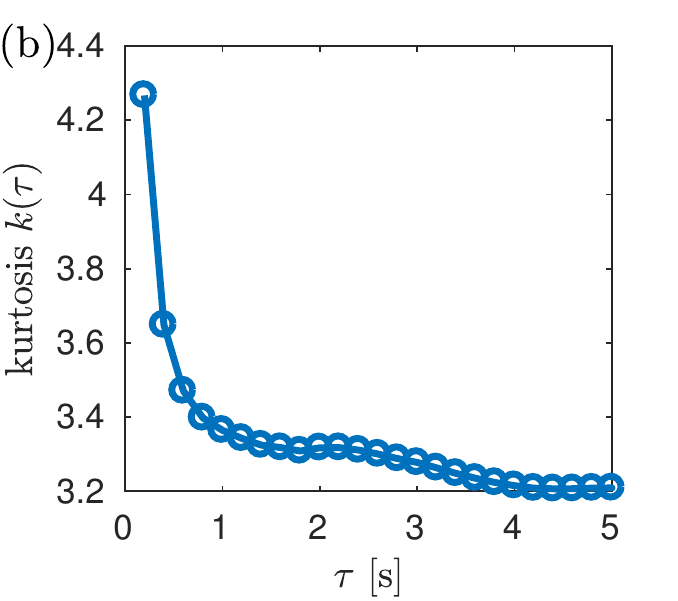}\\
\includegraphics[scale=0.75]{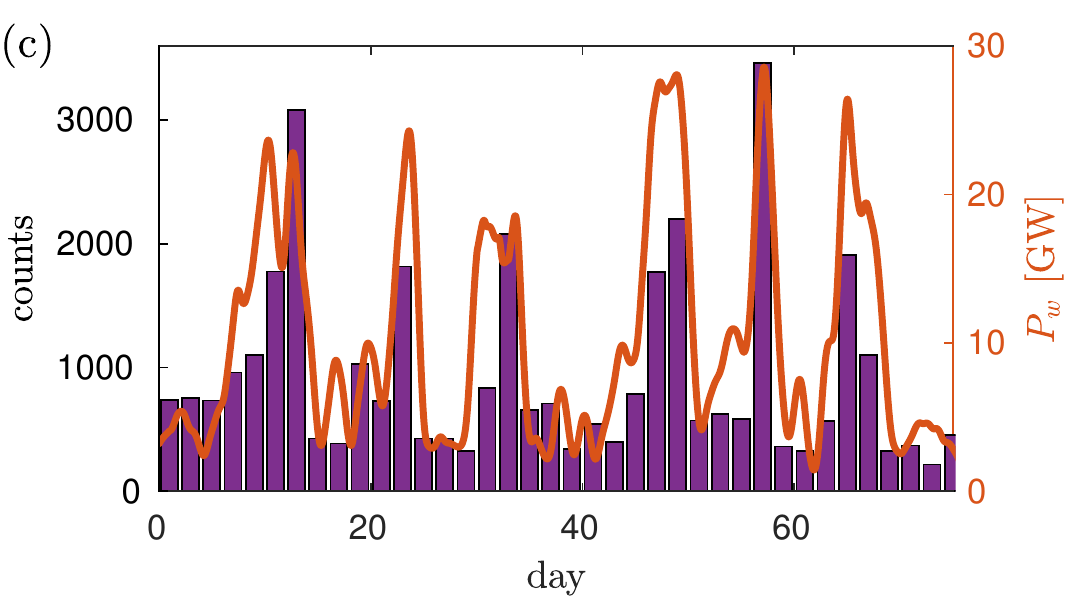}
\caption{{\bf Large {short-term} increments accumulate on days with a high share of wind power fed to the grid.} {\it (a)} Frequency increment distribution $p(\Delta_\tau f)$ for $\tau = 200$ ms. Tails (violet squares) deviate from the normal distribution (gray). { {\it (b)} Kurtosis $k(\tau)$ of $p(\Delta_\tau f)$ as a function of the time lag $\tau$. While for a Gaussian distribution $k=3$, larger kurtosis values correspond to heavier tails. $p(\Delta_\tau f)$ has the highest kurtosis on time scales below one second. {\it (c)} } Left axis and violet boxes: Histogram of occurrences of large increments \mbox{$|\Delta_\tau f| > 2$ mHz} {($\tau = 200$ ms)} binned for two days for the first 70 days of our measurement. Right axis and orange curve: Amount of onshore wind power fed to the grid in Germany. Production data are taken from \cite{entsoe} {and smoothed with moving average of two days}. }
\label{fig:res-incs}
\end{figure}

To evaluate the fluctuations of the grid frequency we focus on probability density functions (PDFs) of increments, $p(\Delta_\tau f)$. As shown in Fig.\ \ref{fig:res-incs} a, the increment distribution $p(\Delta_\tau f)$ is, for a time scale of $\tau = 200$ ms, not Gaussian. Its tails cause strong deviations from the normal distribution. This means that large increments occur much more frequently than expected from a normal distribution.

{The kurtosis $k(\tau) = \langle(\Delta_\tau f - \langle\Delta_\tau f\rangle)^4\rangle / \sigma^4$ measures how heavy-tailed a distribution is. Here, $\sigma$ denotes the standard deviation of $p(\Delta_\tau f)$. The kurtosis takes the value $k=3$ for a Gaussian distribution and increases for more heavy-tailed shapes. We observe that the increment PDFs $p(\Delta_\tau f)$ deform to less heavy-tailed shapes for increasing time lags $\tau$ (Fig.\ \ref{fig:res-incs} b). On time scales below one second, we find the most extreme tails. Even though this effect is similar to turbulent intermittency, can we - at all - relate such short-term fluctuations to wind power injection?}

The grid frequency measurements are, obviously, influenced by many possibly non-Gaussian and/or correlated processes. To investigate a possible dependence of grid fluctuations on wind energy injection, we extract statistical properties of the detrended frequency signal $f(t)$ which we condition on the amount of onshore wind energy $P_w(t)$ that is fed to the European grid in Germany. We use time series provided by the ENTSO-E Transparency Platform \cite{entsoe}, specifically the dataset ``actual generation per production type'' for the country Germany and production type ``onshore wind''. {This data set has a time resolution of 15 minutes. Hence, it does not allow for an analysis of the short-term behavior of the feed-in but still enables us to condition our high-frequency measurements on the amount of wind power fed to the grid.} We focus on the generation in Germany because{, first,} it has, in 2016, by far the highest installed capacity of wind power \cite{windeurope2017} {and, second, the provided data of other countries have an even lower time resolution}. 

We begin with a visual comparison of the increments $\Delta_\tau f$ and $P_w$: In Fig.\ \ref{fig:res-incs} {c}, we show the time instants at which large increments occur as well as the amount of wind power $P_w$ fed to the grid in Germany. {We consider, for the moment, data aggregated for two days.} Large increments $\Delta_\tau f$ coincide with high values of $P_w$. {This is a first indication that wind power feed-in affects grid frequency fluctuations. In the following, we will provide a detailed analysis of the impact of wind power feed-in on frequency increment distributions on different time scales.}

\begin{figure}
\centering
\includegraphics[scale=0.6]{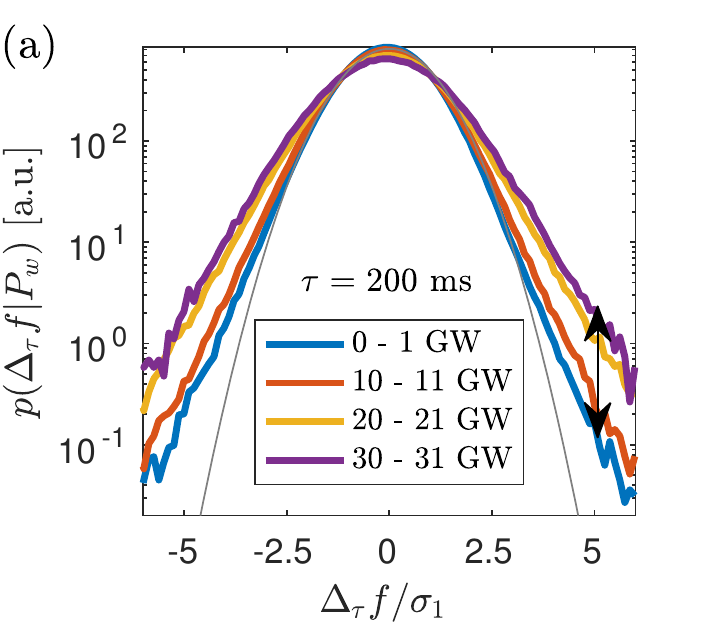}\includegraphics[scale=0.6]{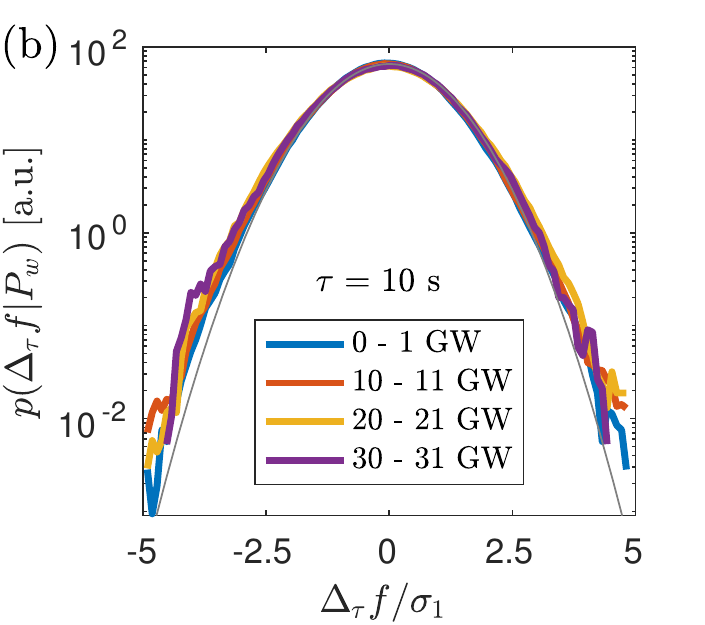}\\
\includegraphics[scale=0.75]{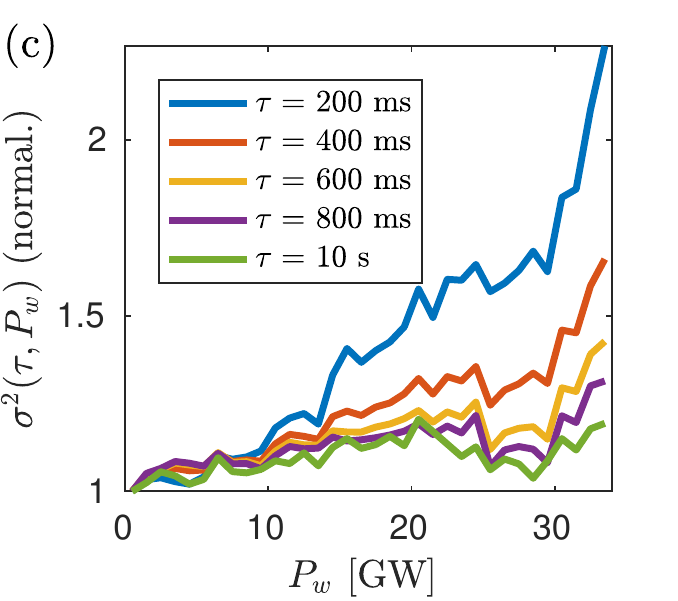}
\caption{{\bf Wind energy feed-in affects the short-term statistics of the power grid frequency.} {\it (a)} PDF of increments $\Delta_\tau f$ on the time scale $\tau = 200$ ms for different intervals of $P_w$ (color-coded) normalized by the standard deviation $\sigma_1$ of the smallest $P_w$ interval. Increasing feed-in $P_w$ broadens the increment distribution resulting in a tenfold higher probability for a $5\sigma_1$-event (black arrow). {\it (b)} Increment PDF for a larger time scale, $\tau = 10$ s. The increments follow the same, almost Gaussian, distribution; independent of the amount of wind energy $P_w$ fed to the grid. {\it (c)} Variances of increment distributions plotted against $P_w$ {for different time scales (color-coded)}. On the {shortest} time scale, $\tau = 200$ ms, the distribution becomes broader with increasing $P_w$. {This effect diminishes with increasing time lags $\tau$.} On the {longest} scale, $\tau = 10$ s, the variance shows no clear trend with $P_w$. For {all} time scales, the variances were normalized with the respective smallest $P_w$ bin.}
\label{fig:res}
\end{figure}

{Instationary stochastic processes are known to potentially produce heavy tails in their probability distributions (e.g. \cite{podobnik2000systems}). Wind turbulence, in particular, shows characteristic turbulent behavior only when wind speed increments are conditioned on the absolute wind speed \cite{morales2012characterization}. Motivated by such findings, we pinpoint the impact of wind power injection on the grid frequency by the analysis of conditioned increment PDFs $p(\Delta_\tau f | P_w)$. Thus, we learn} how likely an increment $\Delta_\tau f$ is if an amount $P_w$ of wind energy is fed to the grid. We show this PDF for a {short} ($\tau = 200$ ms) and a long ($\tau = 10$ s) time scale for different ranges of $P_w$ in Fig. \ref{fig:res} a \& b. First, we observe that on the short scale, the tails deviate from the normal distribution (gray reference curve), whereas the increment PDF is very close to normal on the long scale. Second, for the long time scale, the PDFs are almost identical irrespective of $P_w$. On the short time scale, however, we observe a broadening of the distribution with increasing $P_w$.

{We quantify the time scale dependent impact of the feed-in $P_w$ on the increment PDF by means of width and shape of the conditioned PDFs. In Fig.\ \ref{fig:res} c, we show that the variance }

\begin{align}
\sigma^2(\tau,P_w) := \int (\Delta_\tau f - \langle \Delta_\tau f \rangle_{P_w})^2 p(\Delta_\tau f | P_w)\mathrm{d}\Delta_\tau f
\end{align}
{of the conditioned increment PDF increases with $P_w$ for $\tau = 200$ ms. For increasing time lags $\tau$, this effect quickly diminishes. On time scales of $\tau = 800$ ms and above, the variances show no clear trend with $P_w$.}  \\

{An increased probability of large fluctuations may not only result from an increased variance but also from non-Gaussian, heavy-tailed shapes of the PDFs. To quantify the shape development of $p(\Delta_\tau f|P_w)$ with $P_w$, we use a model which was proposed by Castaing for the characterization of intermittency in turbulence \cite{castaing1990velocity}. }

Castaing uses superimposed Gaussian PDFs with log-normally distributed variances to grasp the tails in intermittent increment PDFs. The standard deviation $\lambda^2$ of the log-normal distribution,

\begin{align}
\lambda^2 = \frac{1}{4}\ln\left(\frac{\langle(\Delta_\tau f - \langle\Delta_\tau f\rangle)^4\rangle}{3\cdot\langle(\Delta_\tau f - \langle\Delta_\tau f\rangle)^2\rangle^2}\right) = \frac{1}{4}\ln\left(\frac{k(\tau)}{3}\right), \label{eq:shape}
\end{align}
governs the shape of the thus obtained PDF $p_c(\Delta_\tau f | P_w)$ and is called {\it shape parameter} \cite{chilla1996multiplicative}. {Due to its close relation to the kurtosis $k(\tau)$, it serves as a measure for the heavy-taildness of the Castaing PDF:} For a Gaussian PDF, $\lambda^2$ is zero. It increases the larger the deviations of the tails from the Gaussian PDF become. In wind speed measurements, we find $\lambda^2$ in the range of 0.2 - 0.3 for increment PDFs on short time scales \cite{morales2012characterization}. 

\begin{figure}
\centering
\includegraphics[scale=0.6]{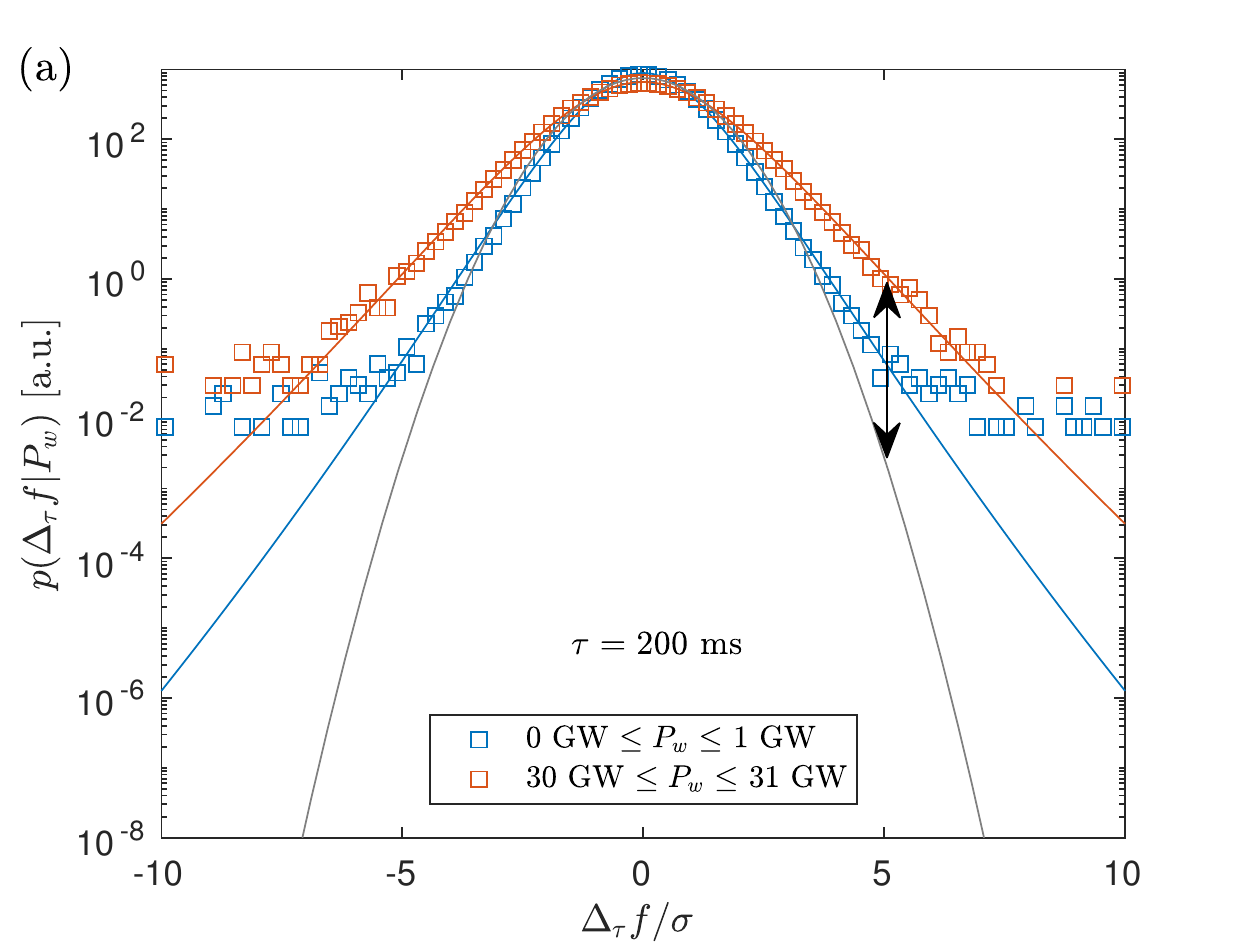}\\
\includegraphics[scale=0.6]{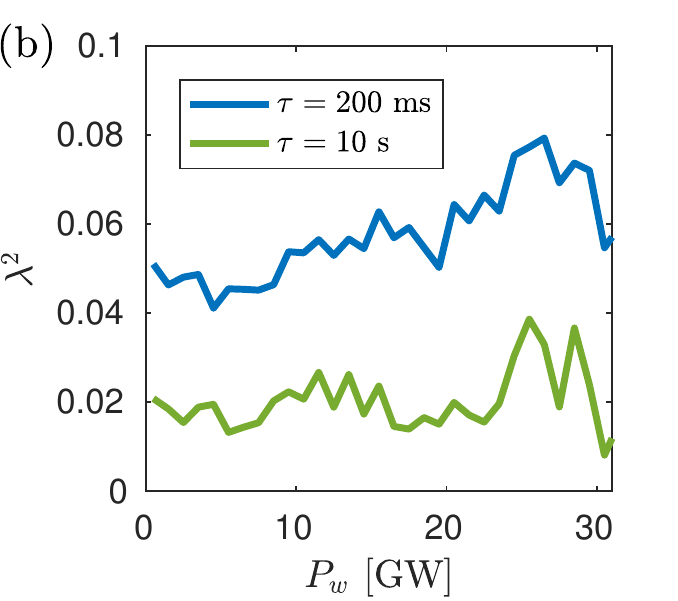}\includegraphics[scale=0.6]{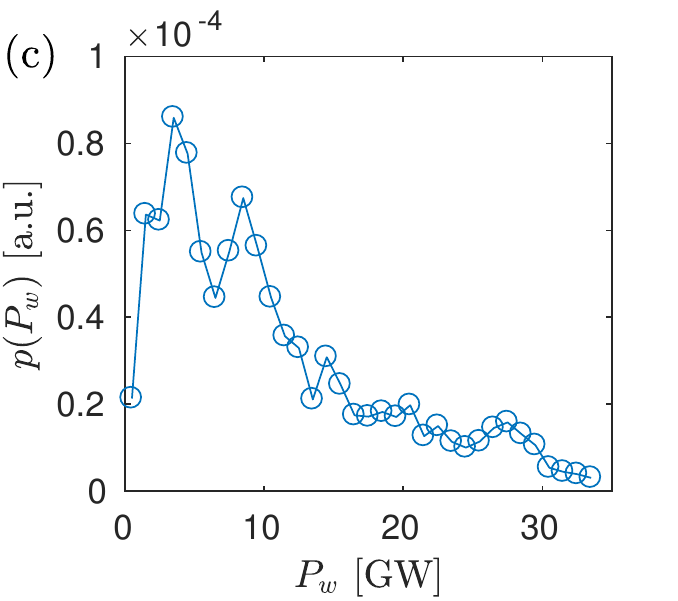}
\caption{{\bf Castaing curves grasp the tails of conditioned increment PDFs.} {\it (a)} Conditioned PDFs $p(\Delta_\tau f | P_w)$ for $\tau = 200$ ms estimated from measurements (squares) and Castaing curves $p_c(\Delta_\tau f | P_w)$ (straight lines) using the shape parameter $\lambda^2$ (Eq.\ \eqref{eq:shape}) for a low (blue) and a high (orange) amount of wind energy feed-in $P_w$. Both PDFs are normalized with the standard deviation $\sigma=0.53$ mHz of the (unconditioned) PDF $p(\Delta_\tau f)$. The Castaing curves emphasize the importance of a correct modeling of the increment PDFs: Within this model, the probability of a 5$\sigma$ event is increased by a factor 900 as compared to a Gaussian model (black arrow). {\it (b)} Shape parameter $\lambda^2$ used to derive $p_c(\Delta_\tau f | P_w)$ {for $\tau = 200$ ms} plotted against $P_w$. {For comparison, we have included the evolution of $\lambda^2$ also for $\tau = 10$ s. The shape parameter shows no clear trend with $P_w$.} {\it (c)} PDF of the wind energy feed-in $p(P_w)$ during our measurements (data available at \cite{entsoe}). }
\label{fig:res-castaing}
\end{figure}

{We calculate $\lambda^2$ from our measurements with Eq.\ \eqref{eq:shape} and follow the steps in \cite{morales2012characterization} to obtain the explicit expression of $p_c(\Delta_\tau f | P_w)$. The results match the data very well as shown in Fig.\ \ref{fig:res-castaing} a, where we} compare the conditioned increment PDFs for a small and a large $P_w$ on the short scale ($\tau = 200$ ms). {With this result we are now able to analyze the change in shape as a function of the wind power $P_w$, see Fig.\ \ref{fig:res-castaing} b. In accordance with Figs.\ \ref{fig:res} a and b, we observe lower $\lambda^2$-values for $\tau = 10$ s than for $\tau = 200$ ms; i.e.\ the PDFs are closer to the Gaussian distribution on the longer time scale. In contrast to the variance (Fig. \ref{fig:res} c), we do not observe a clear trend of $\lambda^2$ with $P_w$. This means that wind energy feed-in mainly broadens the conditioned increment PDF on short scales without much affecting its shape.}

The Castaing parametrizations $p_c(\Delta_\tau f | P_w)$ can be used to estimate the impact of $P_w$ on extreme fluctuations of the frequency: In Fig.\ \ref{fig:res-castaing} a, we compare the probability of a 5$\sigma$-event during high wind energy feed-in $P_w$ to a Gaussian model. We observe a factor 900 between the Castaing model and the Gaussian (black arrow in Fig.\ \ref{fig:res-castaing} a). {Note that this probability factor increases further by many orders of magnitude for larger $\sigma$-events which for the Gaussian statistics are expected to almost never occur - even though we observe them already in our relatively short data set.} This stresses the importance of a correct non-Gaussian modeling of frequency fluctuations in power grids with intermittent feed-in. 

\section{Conclusions and Discussion}
\label{sec:conclusions}

{We have shown that wind power feed-in impacts the power grid frequency on time scales that lie below one second. The time range up to approximately one second is interesting in two aspects: First, it lies in the range of activation of primary frequency control \cite{operation}. This suggests that fluctuations by wind power injection on longer time scales are successfully compensated. Second, in the context of {\it Small Signal Stability Analysis}, one second is approximately the time scale that separates local modes, which affect only a localized subset of nodes in the grid, from so-called interarea modes \cite{zhang2012flexible}. This suggests that the effect we measure is local; a result which is in accordance with our analysis in so far as the used German wind power data are dominated by the northern region of Germany where our frequency measurements were made.}

Power quality is a key challenge for the grid integration of renewable generators \cite{liang2017emerging}. Although the absolute size of the fluctuations we consider is small ($\Delta_\tau f<20$ mHz for $\tau = 200$ ms), a precise knowledge of the fluctuation statistics is essential to correctly estimate the probability of large, possibly critical, increments. {In future power grids with a high share of renewable energy sources, the amount of rotational inertia will be much lower than today. This will lead to faster frequency dynamics with larger amplitudes\cite{ulbig2014impact}. If grid design and control strategies are not properly adapted, such frequency fluctuations may become highly critical for the grid stability \cite{schafer2017escape}. Thus, an explicit expression for increment probabilities is desirable to correctly quantify these risks.}

Our analysis offers a new tool to {quantify} the impact of renewable generation on the frequency increment statistics: The conditioned increment PDFs $p(\Delta_\tau f | P_w)$ are well described by Castaing's parametrization. For a given distribution of the feed-in $p(P_w)$ (Fig.\ \ref{fig:res-castaing} c), the conditioned PDFs may be assumed to follow $p_c(\Delta_\tau f | P_w)$ with, in the simplest model, constant shape parameter (Fig.\ \ref{fig:res-castaing} b) and variance increasing with $P_w$ (Fig.\ \ref{fig:res} c). If shape parameter and variance evolution are inferred from calibration measurements, the increment PDF

\begin{align}
p(\Delta_\tau f) = \int p(P_w)p_c(\Delta_\tau f | P_w) \mathrm{d}P_w
\end{align}
describes the overall impact of wind energy feed-in on the fluctuation characteristics of a given grid {and may be helpful for the design of new control strategies for grids with a high share of renewable sources. 

We want to point out that the non-Gaussian increment statistics $p(\Delta_\tau f)$ may also be fitted with other heavy-tailed distributions, such as q-Gaussians or $\alpha$-stable distributions, which have successfully been applied to single-point PDFs of grid frequency data \cite{schaefer2017nongaussian} as well as to other complex systems like stock markets \cite{podobnik2000scale} or biological systems \cite{ivanov1996scaling}. An important property of such distributions is, besides the stability, the fact that they have diverging moments for wide parameter ranges. We use here the turbulence-like finite-moment approach because, first, we see that power fluctuations are driven by wind turbulence and, second, we observe that frequency increment PDFs are not stable: The sum of two consecutive increments is per definition an increment of a larger scale. However, with increasing time lag, the kurtosis of the increment PDF decreases (Fig.\ \ref{fig:res-incs} b). Hence, the hypothesis of stability is violated.

Independent of the question about the best model of the increment statistics, our main finding is that we observe a broadening of frequency increment PDFs with increasing share of wind power generation. It remains an open question to what extend the shape of the increment PDFs is caused by the turbulent wind statistics \cite{solar} or by other collective effects of interacting grid components. We conclude that a deep understanding of the non-Gaussian fluctuations of renewable energy sources and their interaction with the grid is an important field of further research.}

\acknowledgments
This work was funded by the ministry for science and culture of the German federal state of Lower Saxony (grant no. ZN3045, nieders.\ Vorab). We gratefully acknowledge the technical support from the electronic workshop at Carl von Ossietzky University Oldenburg, especially T. Madena. We would like to thank K. Schmietendorf, M. Anvari, C. Behnken, S. Kettemann, and U. Feudel for valuable discussions.

\bibliography{mybibfile}

\begin{thebibliography}{37}%
\makeatletter
\providecommand \@ifxundefined [1]{%
 \@ifx{#1\undefined}
}%
\providecommand \@ifnum [1]{%
 \ifnum #1\expandafter \@firstoftwo
 \else \expandafter \@secondoftwo
 \fi
}%
\providecommand \@ifx [1]{%
 \ifx #1\expandafter \@firstoftwo
 \else \expandafter \@secondoftwo
 \fi
}%
\providecommand \natexlab [1]{#1}%
\providecommand \enquote  [1]{``#1''}%
\providecommand \bibnamefont  [1]{#1}%
\providecommand \bibfnamefont [1]{#1}%
\providecommand \citenamefont [1]{#1}%
\providecommand \href@noop [0]{\@secondoftwo}%
\providecommand \href [0]{\begingroup \@sanitize@url \@href}%
\providecommand \@href[1]{\@@startlink{#1}\@@href}%
\providecommand \@@href[1]{\endgroup#1\@@endlink}%
\providecommand \@sanitize@url [0]{\catcode `\\12\catcode `\$12\catcode
  `\&12\catcode `\#12\catcode `\^12\catcode `\_12\catcode `\%12\relax}%
\providecommand \@@startlink[1]{}%
\providecommand \@@endlink[0]{}%
\providecommand \url  [0]{\begingroup\@sanitize@url \@url }%
\providecommand \@url [1]{\endgroup\@href {#1}{\urlprefix }}%
\providecommand \urlprefix  [0]{URL }%
\providecommand \Eprint [0]{\href }%
\providecommand \doibase [0]{http://dx.doi.org/}%
\providecommand \selectlanguage [0]{\@gobble}%
\providecommand \bibinfo  [0]{\@secondoftwo}%
\providecommand \bibfield  [0]{\@secondoftwo}%
\providecommand \translation [1]{[#1]}%
\providecommand \BibitemOpen [0]{}%
\providecommand \bibitemStop [0]{}%
\providecommand \bibitemNoStop [0]{.\EOS\space}%
\providecommand \EOS [0]{\spacefactor3000\relax}%
\providecommand \BibitemShut  [1]{\csname bibitem#1\endcsname}%
\let\auto@bib@innerbib\@empty
%</preamble>
\bibitem [{\citenamefont {WindEurope}(2017)}]{windeurope2017}%
  \BibitemOpen
  \bibfield  {author} {\bibinfo {author} {\bibnamefont {WindEurope}},\
  }\href@noop {} {\enquote {\bibinfo {title} {Wind in power - 2016 european
  statistics},}\ } (\bibinfo {year} {2017}),\ \bibinfo {note}
  {\url{https://windeurope.org/wp-content/uploads/files/about-wind/statistics/WindEurope-Annual-Statistics-2016.pdf}}\BibitemShut
  {NoStop}%
\bibitem [{\citenamefont {WindEurope}(2018)}]{dailywind}%
  \BibitemOpen
  \bibfield  {author} {\bibinfo {author} {\bibnamefont {WindEurope}},\
  }\href@noop {} {\enquote {\bibinfo {title} {Daily wind power numbers},}\ }
  (\bibinfo {year} {2018}),\ \bibinfo {note}
  {\url{https://windeurope.org/about-wind/daily-wind/}}\BibitemShut {NoStop}%
\bibitem [{\citenamefont {Kundur}\ \emph {et~al.}(1994)\citenamefont {Kundur},
  \citenamefont {Balu},\ and\ \citenamefont {Lauby}}]{kundur1994power}%
  \BibitemOpen
  \bibfield  {author} {\bibinfo {author} {\bibfnamefont {P.}~\bibnamefont
  {Kundur}}, \bibinfo {author} {\bibfnamefont {N.~J.}\ \bibnamefont {Balu}}, \
  and\ \bibinfo {author} {\bibfnamefont {M.~G.}\ \bibnamefont {Lauby}},\
  }\href@noop {} {\emph {\bibinfo {title} {Power system stability and
  control}}},\ Vol.~\bibinfo {volume} {7}\ (\bibinfo  {publisher} {McGraw-hill
  New York},\ \bibinfo {year} {1994})\BibitemShut {NoStop}%
\bibitem [{ope()}]{operation}%
  \BibitemOpen
  \href@noop {} {\emph {\bibinfo {title} {ENTSO-E Operation Handbook Policy
  1}}},\ \bibinfo {note}
  {\url{https://www.entsoe.eu/fileadmin/user_upload/_library/publications/entsoe/Operation_Handbook/Policy_1_final.pdf/}}\BibitemShut
  {NoStop}%
\bibitem [{\citenamefont {Albadi}\ and\ \citenamefont
  {El-Saadany}(2010)}]{albadi2010overview}%
  \BibitemOpen
  \bibfield  {author} {\bibinfo {author} {\bibfnamefont {M.}~\bibnamefont
  {Albadi}}\ and\ \bibinfo {author} {\bibfnamefont {E.}~\bibnamefont
  {El-Saadany}},\ }\href@noop {} {\bibfield  {journal} {\bibinfo  {journal}
  {Electric Power Systems Research}\ }\textbf {\bibinfo {volume} {80}},\
  \bibinfo {pages} {627} (\bibinfo {year} {2010})}\BibitemShut {NoStop}%
\bibitem [{\citenamefont {Anvari}\ \emph {et~al.}(2016)\citenamefont {Anvari},
  \citenamefont {Lohmann}, \citenamefont {W{\"a}chter}, \citenamefont {Milan},
  \citenamefont {Lorenz}, \citenamefont {Heinemann}, \citenamefont {Tabar},\
  and\ \citenamefont {Peinke}}]{anvari2016short}%
  \BibitemOpen
  \bibfield  {author} {\bibinfo {author} {\bibfnamefont {M.}~\bibnamefont
  {Anvari}}, \bibinfo {author} {\bibfnamefont {G.}~\bibnamefont {Lohmann}},
  \bibinfo {author} {\bibfnamefont {M.}~\bibnamefont {W{\"a}chter}}, \bibinfo
  {author} {\bibfnamefont {P.}~\bibnamefont {Milan}}, \bibinfo {author}
  {\bibfnamefont {E.}~\bibnamefont {Lorenz}}, \bibinfo {author} {\bibfnamefont
  {D.}~\bibnamefont {Heinemann}}, \bibinfo {author} {\bibfnamefont {M.~R.~R.}\
  \bibnamefont {Tabar}}, \ and\ \bibinfo {author} {\bibfnamefont
  {J.}~\bibnamefont {Peinke}},\ }\href@noop {} {\bibfield  {journal} {\bibinfo
  {journal} {New Journal of Physics}\ }\textbf {\bibinfo {volume} {18}},\
  \bibinfo {pages} {063027} (\bibinfo {year} {2016})}\BibitemShut {NoStop}%
\bibitem [{\citenamefont {Milan}\ \emph {et~al.}(2013)\citenamefont {Milan},
  \citenamefont {W{\"a}chter},\ and\ \citenamefont
  {Peinke}}]{milan2013turbulent}%
  \BibitemOpen
  \bibfield  {author} {\bibinfo {author} {\bibfnamefont {P.}~\bibnamefont
  {Milan}}, \bibinfo {author} {\bibfnamefont {M.}~\bibnamefont {W{\"a}chter}},
  \ and\ \bibinfo {author} {\bibfnamefont {J.}~\bibnamefont {Peinke}},\
  }\href@noop {} {\bibfield  {journal} {\bibinfo  {journal} {Physical review
  letters}\ }\textbf {\bibinfo {volume} {110}},\ \bibinfo {pages} {138701}
  (\bibinfo {year} {2013})}\BibitemShut {NoStop}%
\bibitem [{\citenamefont {Boettcher}\ \emph {et~al.}(2003)\citenamefont
  {Boettcher}, \citenamefont {Renner}, \citenamefont {Waldl},\ and\
  \citenamefont {Peinke}}]{boettcher2003statistics}%
  \BibitemOpen
  \bibfield  {author} {\bibinfo {author} {\bibfnamefont {F.}~\bibnamefont
  {Boettcher}}, \bibinfo {author} {\bibfnamefont {C.}~\bibnamefont {Renner}},
  \bibinfo {author} {\bibfnamefont {H.-P.}\ \bibnamefont {Waldl}}, \ and\
  \bibinfo {author} {\bibfnamefont {J.}~\bibnamefont {Peinke}},\ }\href@noop {}
  {\bibfield  {journal} {\bibinfo  {journal} {Boundary-Layer Meteorology}\
  }\textbf {\bibinfo {volume} {108}},\ \bibinfo {pages} {163} (\bibinfo {year}
  {2003})}\BibitemShut {NoStop}%
\bibitem [{\citenamefont {Baile}\ and\ \citenamefont
  {Muzy}(2010)}]{baile2010spatial}%
  \BibitemOpen
  \bibfield  {author} {\bibinfo {author} {\bibfnamefont {R.}~\bibnamefont
  {Baile}}\ and\ \bibinfo {author} {\bibfnamefont {J.-F.}\ \bibnamefont
  {Muzy}},\ }\href@noop {} {\bibfield  {journal} {\bibinfo  {journal} {Physical
  review letters}\ }\textbf {\bibinfo {volume} {105}},\ \bibinfo {pages}
  {254501} (\bibinfo {year} {2010})}\BibitemShut {NoStop}%
\bibitem [{\citenamefont {Calif}\ \emph {et~al.}(2013)\citenamefont {Calif},
  \citenamefont {Schmitt},\ and\ \citenamefont
  {Huang}}]{calif2013multifractal}%
  \BibitemOpen
  \bibfield  {author} {\bibinfo {author} {\bibfnamefont {R.}~\bibnamefont
  {Calif}}, \bibinfo {author} {\bibfnamefont {F.~G.}\ \bibnamefont {Schmitt}},
  \ and\ \bibinfo {author} {\bibfnamefont {Y.}~\bibnamefont {Huang}},\
  }\href@noop {} {\bibfield  {journal} {\bibinfo  {journal} {Physica A:
  Statistical Mechanics and Its Applications}\ }\textbf {\bibinfo {volume}
  {392}},\ \bibinfo {pages} {4106} (\bibinfo {year} {2013})}\BibitemShut
  {NoStop}%
\bibitem [{\citenamefont {Renner}\ \emph {et~al.}(2001)\citenamefont {Renner},
  \citenamefont {Peinke},\ and\ \citenamefont
  {Friedrich}}]{renner2001experimental}%
  \BibitemOpen
  \bibfield  {author} {\bibinfo {author} {\bibfnamefont {C.}~\bibnamefont
  {Renner}}, \bibinfo {author} {\bibfnamefont {J.}~\bibnamefont {Peinke}}, \
  and\ \bibinfo {author} {\bibfnamefont {R.}~\bibnamefont {Friedrich}},\
  }\href@noop {} {\bibfield  {journal} {\bibinfo  {journal} {Journal of Fluid
  Mechanics}\ }\textbf {\bibinfo {volume} {433}},\ \bibinfo {pages} {383}
  (\bibinfo {year} {2001})}\BibitemShut {NoStop}%
\bibitem [{\citenamefont {Frisch}(1995)}]{frisch1995turbulence}%
  \BibitemOpen
  \bibfield  {author} {\bibinfo {author} {\bibfnamefont {U.}~\bibnamefont
  {Frisch}},\ }\href@noop {} {\emph {\bibinfo {title} {Turbulence: the legacy
  of AN Kolmogorov}}}\ (\bibinfo  {publisher} {Cambridge university press},\
  \bibinfo {year} {1995})\BibitemShut {NoStop}%
\bibitem [{\citenamefont {Liang}(2017)}]{liang2017emerging}%
  \BibitemOpen
  \bibfield  {author} {\bibinfo {author} {\bibfnamefont {X.}~\bibnamefont
  {Liang}},\ }\href@noop {} {\bibfield  {journal} {\bibinfo  {journal} {IEEE
  Transactions on Industry Applications}\ }\textbf {\bibinfo {volume} {53}},\
  \bibinfo {pages} {855} (\bibinfo {year} {2017})}\BibitemShut {NoStop}%
\bibitem [{\citenamefont {Filatrella}\ \emph {et~al.}(2008)\citenamefont
  {Filatrella}, \citenamefont {Nielsen},\ and\ \citenamefont
  {Pedersen}}]{filatrella2008analysis}%
  \BibitemOpen
  \bibfield  {author} {\bibinfo {author} {\bibfnamefont {G.}~\bibnamefont
  {Filatrella}}, \bibinfo {author} {\bibfnamefont {A.~H.}\ \bibnamefont
  {Nielsen}}, \ and\ \bibinfo {author} {\bibfnamefont {N.~F.}\ \bibnamefont
  {Pedersen}},\ }\href@noop {} {\bibfield  {journal} {\bibinfo  {journal} {The
  European Physical Journal B-Condensed Matter and Complex Systems}\ }\textbf
  {\bibinfo {volume} {61}},\ \bibinfo {pages} {485} (\bibinfo {year}
  {2008})}\BibitemShut {NoStop}%
\bibitem [{\citenamefont {Schmietendorf}\ \emph {et~al.}(2014)\citenamefont
  {Schmietendorf}, \citenamefont {Peinke}, \citenamefont {Friedrich},\ and\
  \citenamefont {Kamps}}]{schmietendorf2014self}%
  \BibitemOpen
  \bibfield  {author} {\bibinfo {author} {\bibfnamefont {K.}~\bibnamefont
  {Schmietendorf}}, \bibinfo {author} {\bibfnamefont {J.}~\bibnamefont
  {Peinke}}, \bibinfo {author} {\bibfnamefont {R.}~\bibnamefont {Friedrich}}, \
  and\ \bibinfo {author} {\bibfnamefont {O.}~\bibnamefont {Kamps}},\
  }\href@noop {} {\bibfield  {journal} {\bibinfo  {journal} {The European
  Physical Journal Special Topics}\ }\textbf {\bibinfo {volume} {223}},\
  \bibinfo {pages} {2577} (\bibinfo {year} {2014})}\BibitemShut {NoStop}%
\bibitem [{\citenamefont {Rohden}\ \emph {et~al.}(2012)\citenamefont {Rohden},
  \citenamefont {Sorge}, \citenamefont {Timme},\ and\ \citenamefont
  {Witthaut}}]{rohden2012self}%
  \BibitemOpen
  \bibfield  {author} {\bibinfo {author} {\bibfnamefont {M.}~\bibnamefont
  {Rohden}}, \bibinfo {author} {\bibfnamefont {A.}~\bibnamefont {Sorge}},
  \bibinfo {author} {\bibfnamefont {M.}~\bibnamefont {Timme}}, \ and\ \bibinfo
  {author} {\bibfnamefont {D.}~\bibnamefont {Witthaut}},\ }\href@noop {}
  {\bibfield  {journal} {\bibinfo  {journal} {Physical review letters}\
  }\textbf {\bibinfo {volume} {109}},\ \bibinfo {pages} {064101} (\bibinfo
  {year} {2012})}\BibitemShut {NoStop}%
\bibitem [{\citenamefont {Menck}\ \emph {et~al.}(2014)\citenamefont {Menck},
  \citenamefont {Heitzig}, \citenamefont {Kurths},\ and\ \citenamefont
  {Schellnhuber}}]{menck2014dead}%
  \BibitemOpen
  \bibfield  {author} {\bibinfo {author} {\bibfnamefont {P.~J.}\ \bibnamefont
  {Menck}}, \bibinfo {author} {\bibfnamefont {J.}~\bibnamefont {Heitzig}},
  \bibinfo {author} {\bibfnamefont {J.}~\bibnamefont {Kurths}}, \ and\ \bibinfo
  {author} {\bibfnamefont {H.~J.}\ \bibnamefont {Schellnhuber}},\ }\href@noop
  {} {\bibfield  {journal} {\bibinfo  {journal} {Nature communications}\
  }\textbf {\bibinfo {volume} {5}},\ \bibinfo {pages} {3969} (\bibinfo {year}
  {2014})}\BibitemShut {NoStop}%
\bibitem [{\citenamefont {Kettemann}(2016)}]{kettemann2016delocalization}%
  \BibitemOpen
  \bibfield  {author} {\bibinfo {author} {\bibfnamefont {S.}~\bibnamefont
  {Kettemann}},\ }\href@noop {} {\bibfield  {journal} {\bibinfo  {journal}
  {Physical Review E}\ }\textbf {\bibinfo {volume} {94}},\ \bibinfo {pages}
  {062311} (\bibinfo {year} {2016})}\BibitemShut {NoStop}%
\bibitem [{\citenamefont {Sch{\"a}fer}\ \emph {et~al.}(2017)\citenamefont
  {Sch{\"a}fer}, \citenamefont {Matthiae}, \citenamefont {Zhang}, \citenamefont
  {Rohden}, \citenamefont {Timme},\ and\ \citenamefont
  {Witthaut}}]{schafer2017escape}%
  \BibitemOpen
  \bibfield  {author} {\bibinfo {author} {\bibfnamefont {B.}~\bibnamefont
  {Sch{\"a}fer}}, \bibinfo {author} {\bibfnamefont {M.}~\bibnamefont
  {Matthiae}}, \bibinfo {author} {\bibfnamefont {X.}~\bibnamefont {Zhang}},
  \bibinfo {author} {\bibfnamefont {M.}~\bibnamefont {Rohden}}, \bibinfo
  {author} {\bibfnamefont {M.}~\bibnamefont {Timme}}, \ and\ \bibinfo {author}
  {\bibfnamefont {D.}~\bibnamefont {Witthaut}},\ }\href@noop {} {\bibfield
  {journal} {\bibinfo  {journal} {Physical Review E}\ }\textbf {\bibinfo
  {volume} {95}},\ \bibinfo {pages} {060203} (\bibinfo {year}
  {2017})}\BibitemShut {NoStop}%
\bibitem [{\citenamefont {Schmietendorf}\ \emph {et~al.}(2017)\citenamefont
  {Schmietendorf}, \citenamefont {Peinke},\ and\ \citenamefont
  {Kamps}}]{schmietendorf2017impact}%
  \BibitemOpen
  \bibfield  {author} {\bibinfo {author} {\bibfnamefont {K.}~\bibnamefont
  {Schmietendorf}}, \bibinfo {author} {\bibfnamefont {J.}~\bibnamefont
  {Peinke}}, \ and\ \bibinfo {author} {\bibfnamefont {O.}~\bibnamefont
  {Kamps}},\ }\href@noop {} {\bibfield  {journal} {\bibinfo  {journal} {The
  European Physical Journal B}\ }\textbf {\bibinfo {volume} {90}},\ \bibinfo
  {pages} {222} (\bibinfo {year} {2017})}\BibitemShut {NoStop}%
\bibitem [{\citenamefont {Auer}\ \emph {et~al.}(2017)\citenamefont {Auer},
  \citenamefont {Hellmann}, \citenamefont {Krause},\ and\ \citenamefont
  {Kurths}}]{auer2017stability}%
  \BibitemOpen
  \bibfield  {author} {\bibinfo {author} {\bibfnamefont {S.}~\bibnamefont
  {Auer}}, \bibinfo {author} {\bibfnamefont {F.}~\bibnamefont {Hellmann}},
  \bibinfo {author} {\bibfnamefont {M.}~\bibnamefont {Krause}}, \ and\ \bibinfo
  {author} {\bibfnamefont {J.}~\bibnamefont {Kurths}},\ }\href@noop {}
  {\bibfield  {journal} {\bibinfo  {journal} {Chaos: An Interdisciplinary
  Journal of Nonlinear Science}\ }\textbf {\bibinfo {volume} {27}},\ \bibinfo
  {pages} {127003} (\bibinfo {year} {2017})}\BibitemShut {NoStop}%
\bibitem [{\citenamefont {Sch{\"a}fer}\ \emph {et~al.}(2018)\citenamefont
  {Sch{\"a}fer}, \citenamefont {Beck}, \citenamefont {Aihara}, \citenamefont
  {Witthaut},\ and\ \citenamefont {Timme}}]{schaefer2017nongaussian}%
  \BibitemOpen
  \bibfield  {author} {\bibinfo {author} {\bibfnamefont {B.}~\bibnamefont
  {Sch{\"a}fer}}, \bibinfo {author} {\bibfnamefont {C.}~\bibnamefont {Beck}},
  \bibinfo {author} {\bibfnamefont {K.}~\bibnamefont {Aihara}}, \bibinfo
  {author} {\bibfnamefont {D.}~\bibnamefont {Witthaut}}, \ and\ \bibinfo
  {author} {\bibfnamefont {M.}~\bibnamefont {Timme}},\ }\href@noop {}
  {\bibfield  {journal} {\bibinfo  {journal} {Nature Energy}\ ,\ \bibinfo
  {pages} {1}} (\bibinfo {year} {2018})}\BibitemShut {NoStop}%
\bibitem [{\citenamefont {Muzy}\ \emph {et~al.}(2010)\citenamefont {Muzy},
  \citenamefont {Ba{\"\i}le},\ and\ \citenamefont
  {Poggi}}]{muzy2010intermittency}%
  \BibitemOpen
  \bibfield  {author} {\bibinfo {author} {\bibfnamefont {J.-F.}\ \bibnamefont
  {Muzy}}, \bibinfo {author} {\bibfnamefont {R.}~\bibnamefont {Ba{\"\i}le}}, \
  and\ \bibinfo {author} {\bibfnamefont {P.}~\bibnamefont {Poggi}},\
  }\href@noop {} {\bibfield  {journal} {\bibinfo  {journal} {Physical Review
  E}\ }\textbf {\bibinfo {volume} {81}},\ \bibinfo {pages} {056308} (\bibinfo
  {year} {2010})}\BibitemShut {NoStop}%
\bibitem [{\citenamefont {Kamps}(2014)}]{kamps2014characterizing}%
  \BibitemOpen
  \bibfield  {author} {\bibinfo {author} {\bibfnamefont {O.}~\bibnamefont
  {Kamps}},\ }\href@noop {} {\bibfield  {journal} {\bibinfo  {journal} {Wind
  Energy-Impact of Turbulence}\ ,\ \bibinfo {pages} {67}} (\bibinfo {year}
  {2014})}\BibitemShut {NoStop}%
\bibitem [{dow()}]{downloadData}%
  \BibitemOpen
  \href@noop {} {}\bibinfo {note}
  {\url{http://www.uni-oldenburg.de/fileadmin/user_upload/physik/ag/twist/Forschung/Daten/AnvariEtAl2016_ExampleData_WindPowerAndIrradianceData.zip}}\BibitemShut
  {NoStop}%
\bibitem [{\citenamefont {Lauby}\ \emph {et~al.}(2014)\citenamefont {Lauby},
  \citenamefont {Bian}, \citenamefont {Ekisheva},\ and\ \citenamefont
  {Varghese}}]{lauby2014frequency}%
  \BibitemOpen
  \bibfield  {author} {\bibinfo {author} {\bibfnamefont {M.}~\bibnamefont
  {Lauby}}, \bibinfo {author} {\bibfnamefont {J.}~\bibnamefont {Bian}},
  \bibinfo {author} {\bibfnamefont {S.}~\bibnamefont {Ekisheva}}, \ and\
  \bibinfo {author} {\bibfnamefont {M.}~\bibnamefont {Varghese}},\ }in\
  \href@noop {} {\emph {\bibinfo {booktitle} {North American Power Symposium
  (NAPS), 2014}}}\ (\bibinfo {organization} {IEEE},\ \bibinfo {year} {2014})\
  pp.\ \bibinfo {pages} {1--5}\BibitemShut {NoStop}%
\bibitem [{\citenamefont {Boashash}(1992)}]{boashash1992estimating}%
  \BibitemOpen
  \bibfield  {author} {\bibinfo {author} {\bibfnamefont {B.}~\bibnamefont
  {Boashash}},\ }\href@noop {} {\bibfield  {journal} {\bibinfo  {journal}
  {Proceedings of the IEEE}\ }\textbf {\bibinfo {volume} {80}},\ \bibinfo
  {pages} {520} (\bibinfo {year} {1992})}\BibitemShut {NoStop}%
\bibitem [{\citenamefont {ENTSO-E}()}]{entsoe}%
  \BibitemOpen
  \bibfield  {author} {\bibinfo {author} {\bibnamefont {ENTSO-E}},\ }\href@noop
  {} {\emph {\bibinfo {title} {Transparency Platform}}},\ \bibinfo {note}
  {\url{https://transparency.entsoe.eu/}}\BibitemShut {NoStop}%
\bibitem [{\citenamefont {Podobnik}\ \emph
  {et~al.}(2000{\natexlab{a}})\citenamefont {Podobnik}, \citenamefont {Ivanov},
  \citenamefont {Lee}, \citenamefont {Chessa},\ and\ \citenamefont
  {Stanley}}]{podobnik2000systems}%
  \BibitemOpen
  \bibfield  {author} {\bibinfo {author} {\bibfnamefont {B.}~\bibnamefont
  {Podobnik}}, \bibinfo {author} {\bibfnamefont {P.~C.}\ \bibnamefont
  {Ivanov}}, \bibinfo {author} {\bibfnamefont {Y.}~\bibnamefont {Lee}},
  \bibinfo {author} {\bibfnamefont {A.}~\bibnamefont {Chessa}}, \ and\ \bibinfo
  {author} {\bibfnamefont {H.~E.}\ \bibnamefont {Stanley}},\ }\href@noop {}
  {\bibfield  {journal} {\bibinfo  {journal} {EPL (Europhysics Letters)}\
  }\textbf {\bibinfo {volume} {50}},\ \bibinfo {pages} {711} (\bibinfo {year}
  {2000}{\natexlab{a}})}\BibitemShut {NoStop}%
\bibitem [{\citenamefont {Morales}\ \emph {et~al.}(2012)\citenamefont
  {Morales}, \citenamefont {W{\"a}chter},\ and\ \citenamefont
  {Peinke}}]{morales2012characterization}%
  \BibitemOpen
  \bibfield  {author} {\bibinfo {author} {\bibfnamefont {A.}~\bibnamefont
  {Morales}}, \bibinfo {author} {\bibfnamefont {M.}~\bibnamefont
  {W{\"a}chter}}, \ and\ \bibinfo {author} {\bibfnamefont {J.}~\bibnamefont
  {Peinke}},\ }\href@noop {} {\bibfield  {journal} {\bibinfo  {journal} {Wind
  Energy}\ }\textbf {\bibinfo {volume} {15}},\ \bibinfo {pages} {391} (\bibinfo
  {year} {2012})}\BibitemShut {NoStop}%
\bibitem [{\citenamefont {Castaing}\ \emph {et~al.}(1990)\citenamefont
  {Castaing}, \citenamefont {Gagne},\ and\ \citenamefont
  {Hopfinger}}]{castaing1990velocity}%
  \BibitemOpen
  \bibfield  {author} {\bibinfo {author} {\bibfnamefont {B.}~\bibnamefont
  {Castaing}}, \bibinfo {author} {\bibfnamefont {Y.}~\bibnamefont {Gagne}}, \
  and\ \bibinfo {author} {\bibfnamefont {E.}~\bibnamefont {Hopfinger}},\
  }\href@noop {} {\bibfield  {journal} {\bibinfo  {journal} {Physica D:
  Nonlinear Phenomena}\ }\textbf {\bibinfo {volume} {46}},\ \bibinfo {pages}
  {177} (\bibinfo {year} {1990})}\BibitemShut {NoStop}%
\bibitem [{\citenamefont {Chilla}\ \emph {et~al.}(1996)\citenamefont {Chilla},
  \citenamefont {Peinke},\ and\ \citenamefont
  {Castaing}}]{chilla1996multiplicative}%
  \BibitemOpen
  \bibfield  {author} {\bibinfo {author} {\bibfnamefont {F.}~\bibnamefont
  {Chilla}}, \bibinfo {author} {\bibfnamefont {J.}~\bibnamefont {Peinke}}, \
  and\ \bibinfo {author} {\bibfnamefont {B.}~\bibnamefont {Castaing}},\
  }\href@noop {} {\bibfield  {journal} {\bibinfo  {journal} {Journal de
  Physique II}\ }\textbf {\bibinfo {volume} {6}},\ \bibinfo {pages} {455}
  (\bibinfo {year} {1996})}\BibitemShut {NoStop}%
\bibitem [{\citenamefont {Zhang}\ \emph {et~al.}(2012)\citenamefont {Zhang},
  \citenamefont {Rehtanz},\ and\ \citenamefont {Pal}}]{zhang2012flexible}%
  \BibitemOpen
  \bibfield  {author} {\bibinfo {author} {\bibfnamefont {X.-P.}\ \bibnamefont
  {Zhang}}, \bibinfo {author} {\bibfnamefont {C.}~\bibnamefont {Rehtanz}}, \
  and\ \bibinfo {author} {\bibfnamefont {B.}~\bibnamefont {Pal}},\ }\href@noop
  {} {\emph {\bibinfo {title} {Flexible AC transmission systems: modelling and
  control}}}\ (\bibinfo  {publisher} {Springer Science \& Business Media},\
  \bibinfo {year} {2012})\BibitemShut {NoStop}%
\bibitem [{\citenamefont {Ulbig}\ \emph {et~al.}(2014)\citenamefont {Ulbig},
  \citenamefont {Borsche},\ and\ \citenamefont {Andersson}}]{ulbig2014impact}%
  \BibitemOpen
  \bibfield  {author} {\bibinfo {author} {\bibfnamefont {A.}~\bibnamefont
  {Ulbig}}, \bibinfo {author} {\bibfnamefont {T.~S.}\ \bibnamefont {Borsche}},
  \ and\ \bibinfo {author} {\bibfnamefont {G.}~\bibnamefont {Andersson}},\
  }\href@noop {} {\bibfield  {journal} {\bibinfo  {journal} {IFAC Proceedings
  Volumes}\ }\textbf {\bibinfo {volume} {47}},\ \bibinfo {pages} {7290}
  (\bibinfo {year} {2014})}\BibitemShut {NoStop}%
\bibitem [{\citenamefont {Podobnik}\ \emph
  {et~al.}(2000{\natexlab{b}})\citenamefont {Podobnik}, \citenamefont {Ivanov},
  \citenamefont {Lee},\ and\ \citenamefont {Stanley}}]{podobnik2000scale}%
  \BibitemOpen
  \bibfield  {author} {\bibinfo {author} {\bibfnamefont {B.}~\bibnamefont
  {Podobnik}}, \bibinfo {author} {\bibfnamefont {P.~C.}\ \bibnamefont
  {Ivanov}}, \bibinfo {author} {\bibfnamefont {Y.}~\bibnamefont {Lee}}, \ and\
  \bibinfo {author} {\bibfnamefont {H.~E.}\ \bibnamefont {Stanley}},\
  }\href@noop {} {\bibfield  {journal} {\bibinfo  {journal} {EPL (Europhysics
  Letters)}\ }\textbf {\bibinfo {volume} {52}},\ \bibinfo {pages} {491}
  (\bibinfo {year} {2000}{\natexlab{b}})}\BibitemShut {NoStop}%
\bibitem [{\citenamefont {Ivanov}\ \emph {et~al.}(1996)\citenamefont {Ivanov},
  \citenamefont {Rosenblum}, \citenamefont {Peng}, \citenamefont {Mietus},
  \citenamefont {Havlin}, \citenamefont {Stanley},\ and\ \citenamefont
  {Goldberger}}]{ivanov1996scaling}%
  \BibitemOpen
  \bibfield  {author} {\bibinfo {author} {\bibfnamefont {P.~C.}\ \bibnamefont
  {Ivanov}}, \bibinfo {author} {\bibfnamefont {M.~G.}\ \bibnamefont
  {Rosenblum}}, \bibinfo {author} {\bibfnamefont {C.-K.}\ \bibnamefont {Peng}},
  \bibinfo {author} {\bibfnamefont {J.}~\bibnamefont {Mietus}}, \bibinfo
  {author} {\bibfnamefont {S.}~\bibnamefont {Havlin}}, \bibinfo {author}
  {\bibfnamefont {H.~E.}\ \bibnamefont {Stanley}}, \ and\ \bibinfo {author}
  {\bibfnamefont {A.~L.}\ \bibnamefont {Goldberger}},\ }\href@noop {}
  {\bibfield  {journal} {\bibinfo  {journal} {Nature}\ }\textbf {\bibinfo
  {volume} {383}},\ \bibinfo {pages} {323} (\bibinfo {year}
  {1996})}\BibitemShut {NoStop}%
\bibitem [{sol()}]{solar}%
  \BibitemOpen
  \href@noop {} {}\bibinfo {note} {In principle, we expect also photovoltaic
  feed-in to cause similar frequency fluctuations. However, we leave this as a
  question for further research because, first, during our winter time
  measurements only small amounts of PV energy were generated and, second,
  intermittency in PV power time series depends not only on the amount of
  produced energy but also on cloud structures. Hence, a similar analysis for
  PV requires additional data sets.}\BibitemShut {Stop}%
\end{thebibliography}%

\end{document}